\documentclass[fleqn,10pt]{wlscirep}
\usepackage{mathtools}
\usepackage[utf8]{inputenc}
\usepackage[T1]{fontenc}

\usepackage{setspace}
\usepackage{graphicx}
\usepackage{wasysym} 
\usepackage{graphicx}% Include figure files
\usepackage{dcolumn}% Align table columns on decimal point
\usepackage{bm}% bold math
\usepackage{color}

\usepackage[english]{babel}
\usepackage[utf8]{inputenc}

\title{Information parity on cortical functional brain networks increases under psychedelic influences}
\author[1,2,3*]{Aline Viol}
\author[4,5]{Gandhi M. Viswanathan}
\author[1,6]{Oleksandra Soldatkina}
\author[7]{Fernanda Palhano-Fontes}
\author[7]{Heloisa Onias}
\author[7]{Draulio de Araujo}
\author[8]{Philipp H\"ovel}

\affil[1]{\small Scuola Internazionale Superiore di Studi Avanzati - SISSA, Cognitive Neuroscience, Via Bonomea 265, 34136, Trieste, Italy}
 \affil[2] {\small Institute of Theoretical Physics, Technische Universit\"at Berlin, Hardenbergstra\ss{}e 36, 10623 Berlin, Germany}
\affil[3]{\small Bernstein Center for Computational Neuroscience Berlin, Humboldt-Universit{\"a}t zu Berlin, Philippstra{\ss}e 13, 10115 Berlin, Germany}
\affil[4]{\small Department of Theoretical and Experimental Physics, Federal University of Rio Grande do Norte, Natal--RN, 59078-970,
Brazil}
\affil[5]{\small National Institute of Science and Technology of Complex Systems,
Federal University of Rio Grande do Norte, Natal--RN, 59078-970,
Brazil}
\affil[6]{\small Barcelona Supercomputing Center, Barcelona,08034, Spain.}

\affil[7]{\small Brain Institute, Federal University of Rio Grande do Norte,
Natal--RN, 59078-970, Brazil}
\affil[8]{\small School of Mathematical Sciences, University College Cork, Western Road T12 XF62, Cork, Ireland}

\affil[*]{\small{aline.viol@sissa.it}}

\begin{abstract}

The physical basis of consciousness is one of the most intriguing open questions that contemporary science aims to solve. By approaching the brain as an interactive information system, complex network theory has greatly contributed to understand brain process in different states of mind. 
We study an non-ordinary state of mind by comparing resting-state functional brain networks of individuals in two different conditions: before and after the ingestion of the psychedelic brew Ayahuasca.
In order to quantify the functional, statistical symmetries between brain region connectivity, we calculate the pairwise information parity of the functional brain networks. Unlike most of usual network quantification, which is done on a local or global scale, information parity quantifies the pairwise statistical similarities considering the entire network structure.
We detect an increase in the average information parity on brain networks of individuals under psychedelic influences. Notably, the information parity between regions from the limbic system and frontal cortex is consistently higher for all the individuals while under the psychedelic influence. 
\end{abstract}

\begin{document}
\maketitle

\thispagestyle{empty}

\section*{Introduction}

Understanding the relationship between mental states and brain activity is one of the most intricate challenges of neuroscience.
Finding distinguishing features of the human brain mechanisms considering different states of mind seems to be a reliable approach toward unraveling the neural correlate of consciousness. In fact, several studies using neuroimaging data have shown consistent correlations between mental states and functional brain networks in humans  \cite{TAG14,ONIAS14,VIO17a,VAR20,CAR12,DAN21,LEW12}. Emergent behaviors on networked systems are determined by their topological structure. For instance, topological features such as quenched disorders and modularity are known to influence key aspects of criticality \cite{GUT21,MATA15,SAN19}, what have been related to brain dynamics \cite{COC17}. 
In this paper, we intend to shed light on the following question: Is some form of symmetry affected in the functional brain networks in different states of consciousness?

To address this question, we evaluate the information parity on functional brain networks of individuals in the ordinary state of mind and under the effects of a potent psychedelic substance. 
The information parity is a measurement that estimates the similarity of influences on a pair of nodes considering their relative location in the network. 
A node in a network influences and is influenced not only by its first neighbors but by the network structure as a whole. We assess the ``topological profile" of a node in terms of the whole network structure considered from the viewpoint of the referred node. This means a mapping of which nodes are in the nearest neighborhood, in the next-nearest neighborhood, and so on. We refer to the reference~\cite{VIOL19} for further elucidation about topological profiles. 
For instance, in the case of networks in which the topology completely defines the node roles, nodes with the same topological profile would have exactly the same role. 
Information parity estimates the common information collected by a pair of nodes, while also taking into account their individual statistics \cite{VIOL21}.
In simple words, information parity quantifies how much information one can have about a node by knowing the topological profile of another node. 
From the information theory perspective, information parity is the excerpt of mutual information responsible for quantifying the statistical symmetry.   
Moreover, being a local measure that encapsulates global information, information parity can help map network resilience and structural hierarchy.

We argue whether the information parity on functional brain networks would change in individuals under influence of psychedelics. We evaluated functional magnetic resonance imaging (fMRI) from volunteers who take a psychedelic brew, called Ayahuasca, from Amazonian indigenous cultures. The main psychoactive components of Ayahuasca are the N,N - Dimethyltryptamine (DMT), and Monoamine oxidase inhibitors (MOIs). DMT is a serotonergic psychedelic substance that is quickly degraded in the human body. Hence, the presence of the MOIs in the decoction avoids its degradation allowing the DMT to cross the blood-brain barrier. Although Ayahuasca, as well as other psychedelic substances, causes considerable changes in perception and cognition \cite{SAN16,WEISS21}, characterizing the changes in the functional brain networks is not a straightforward task, mainly due to the wide variability among individuals. 
In previous works, we detected increasing entropy on function networks features of individuals after Ayahuasca intake \cite{VIO17a,VIOL19,FEL21}. 
Here, we report a persistent increase in the average information parity for all individuals under the psychedelic influence. The latter result suggests that the increase of entropy may be coordinated with an increase in redundancy. 
%the increase of pairwise statistical similarities between brain regions. 
We also detect that, despite great variability in the patterns of information parity among the individuals, all of the individuals have an increased information parity between the frontal cortex and regions of the limbic system. It has been suggested that frontal cortex regions are involve in top-down modulation of external attention and emotions \cite{BUS07,HOP2000,OCH02} while the limbic regions as hippocampus and amygdala are involved in internal process as memory an emotions \cite{BAT18,CAR11}. Next, we briefly describe the methods and, after reporting the results, we discuss possible interpretations in the context of the quantitative neuroscience of psychedelics.

 \section*{Methods}

\subsection*{Neuroimaging data}

The data used in the present analysis is the same of an previously published studies \cite{VIO17a,VIOL19}. The reader can find carefully described details of acquisition in the reference \cite{VIO17a}. In the following, we briefly summarize  the essential information. 
The data consist of 7 right-handed adults volunteers ($\approx$ 31.3 years old, from 24 to 47 years) in absence of medication influence at least 3 months prior to the acquisition sessions. They were in abstinence from alcohol, caffeine and nicotine and had attested psychiatric or neurological disorders (assessed by DSM-IV structured interview). The acquisition protocol was approved by the Ethics and Research Committee of the University of São Paulo at Ribeirão Preto (process number 14672/2006)\cite{ARAU12}. 
All the volunteers were submitted twice to fMRI scanning: one before and the other 40 minutes after Ayahuasca intake. In both scanning sections, they were requested to remain in awake resting state, that is, lying down, with eyes closed, with the mind free for wandering.
Each volunteer ingested 120–200 mL (2.2 mL/kg of body weight) of Ayahuasca. The brew contained 0.8 mg/mL of DMT and 0.21 mg/mL of harmine. Harmaline was not detected  at the threshold of 0.02 mg/mL via the chromatography analysis. The data were preprocessed according to the following steps: slice-timing and head motion correction, and spatial smoothing (Gaussian kernel, FWHM = 5 mm). The samples were spatially normalized to the Montreal Neurologic Institute (MNI152). It was calculated 9 regressors using a General Linear Model (GLM): 6 regressors to movement correction, 1 to white matter signal, 1 to cerebrospinal fluid and 1 to global signal. 

\subsection*{Functional brain networks}

In order to define the networks, we segmented the 3D brain images into 110 brain regions according to the Harvard-Oxford cortical and subcortical structural atlas. Due to technical limitations, 6 regions had to be excluded from the analysis. 
We extracted one time series from each region by computing the average within that region and applied a band-pass filter ($ \approx 0.03 - 0.07$ Hz) using the maximum overlap wavelet transform (MODWT). In our approach, the brain regions define the network nodes, and links are defined by the Pearson correlation between the wavelet coefficients, yielding a $104 \times 104$ correlation matrix. 
We obtain adjacency matrices $A$ representing unweighted undirected networks by applying thresholds on the absolute value of the correlation matrices. That is if the values are larger than the threshold, the link $A_{i,j}$ is defined to be 1,
otherwise, it is defined to be 0. Since there is no solid method to estimate a priori the most appropriate threshold, to ensure the robustness of our analysis we evaluate a range of thresholds generating a set of networks with different densities for each sample. We used the same procedure used in references \cite{VIO17a,VIOL19}. Considering that information parity, as most of the network measurements, is sensible to network density, we were careful in comparing networks before and after Ayahuasca intake with the same densities. 

\subsection*{Information parity}

Given a network  
$G(V, E)$, in which $V$ is a set of $N$ nodes and $E$ is the set of their undirected and unweighted links. The network is represented by the adjacency matrix $A$ defined $1$ for a pair $(i,j)$ of connected nodes and $0$, otherwise. The information parity \cite{VIOL21} between a pair of nodes ($i,j$) is defined as:
\begin{equation}
    I_{i,j}=\sum_{r=1}^{r_{max}}p_{i,j}(r)\log{\frac{p_{i,j}(r)}{p_{i}(r)p_{j}(r)}} ~,
\end{equation}
in which $r$ are integer in the interval $1\leq r\leq r_{\textrm{max}}$, in with $r_{\textrm{max}}$ is the maximum neighborhood radius of the nodes \cite{VIOL19,VIOL21}. The probability $p_i(r)$ of find a node at $r$ links of distance from the node $i$ and the probability $p_{i,j}(r)$ of find a node at the same $r$ distance from the nodes $i$ and $j$ are respectively defined as following:
\begin{equation} \label{probi}
    p_{i}(r)= \frac{1}{N-1}\sum_{\substack{ k \in V \\ k\neq i}} \delta(D_{ik},r) ~,
    \end{equation}
    and, 
    \begin{equation} \label{probij}
    p_{ij}(r)= \frac{1}{N-2}\sum_{\substack{ k \in V \\ k\neq i,j}} \delta(D_{ik},r) \, \delta(D_{jk},r) ~,
\end{equation}
where $\delta(\cdot,\cdot)$ denotes the Kronecker and $D_{ik}$ the geodesic distance between the node $i$ and the node $k$ \cite{VIOL21}. 

\section*{Results}
\begin{figure}[ht]
\centering
a) \includegraphics[width=0.7\linewidth]{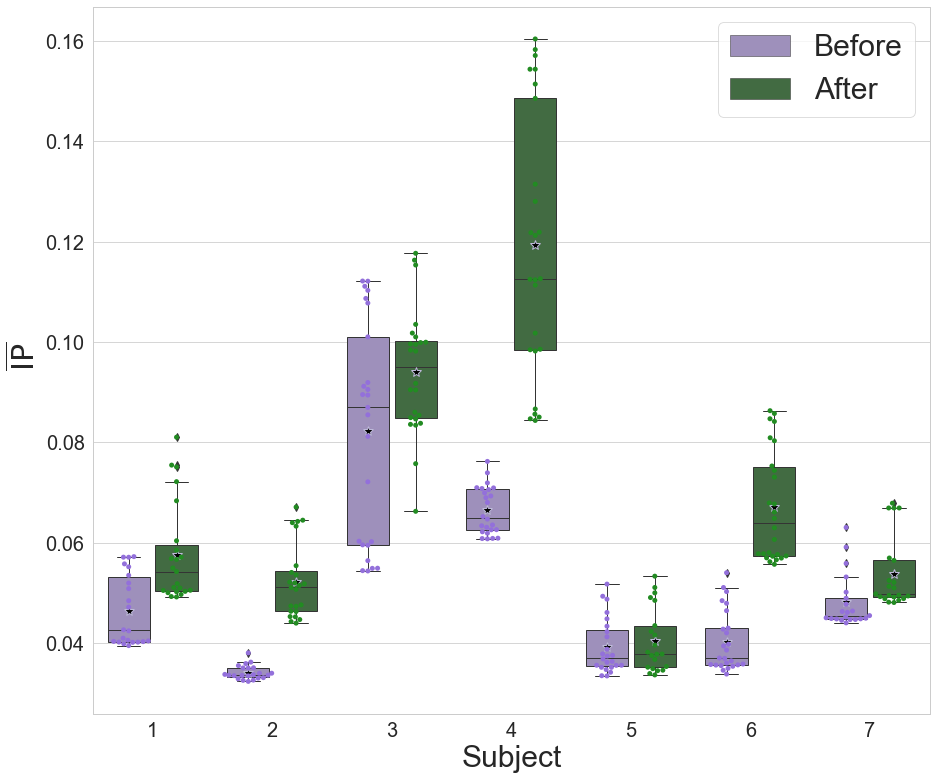}

b) \includegraphics[width=0.7\linewidth]{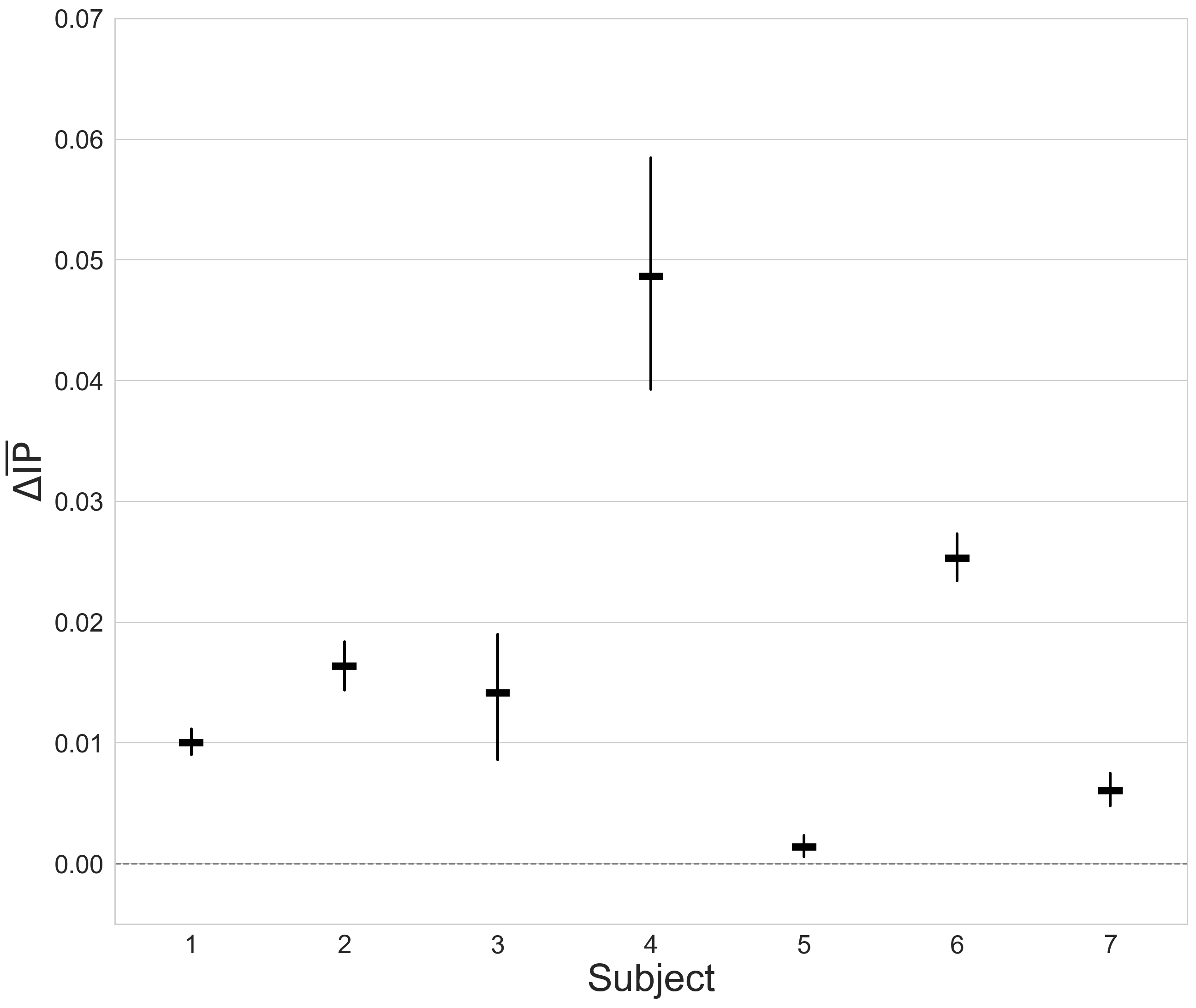}
\caption{Information parity increases after Ayahuasca intake. The boxplot in a) shows the average information parity before in purple and after Ayahuasca intake in green. Each depicted point represents the average over all pairs of nodes for networks with different densities. The stars mark the mean value across all networks from each sample. The plots in b) depicts the average of the information parity divergence ($\rm IP_{after}- IP_{before})$ between networks with the same density. Note that the information parity increases for all the subjects. }
\label{fig:main_result}

\end{figure}

In this section, we report the detected changes when we compared the functional brain networks before and after Ayahuasca intake. 
Figure~\ref{fig:main_result}~a) shows the average information parity for each subject before (purple) and after Ayahuasca intake (green). Each boxplot depicts the information parity of a set of networks with different densities, generated from the same correlation matrix. Each dot represents the average information parity for one network. Note that, on average, the network's information parity increases for all subjects after Ayahuasca intake.
Figure~\ref{fig:main_result}~b) shows the difference between after and before Ayahuasca intake comparing networks with same density. The comparisons for all subjects resulted in positive differences independent of the network density. 

\begin{figure}[ht]
\centering
\includegraphics[width=0.49\linewidth]{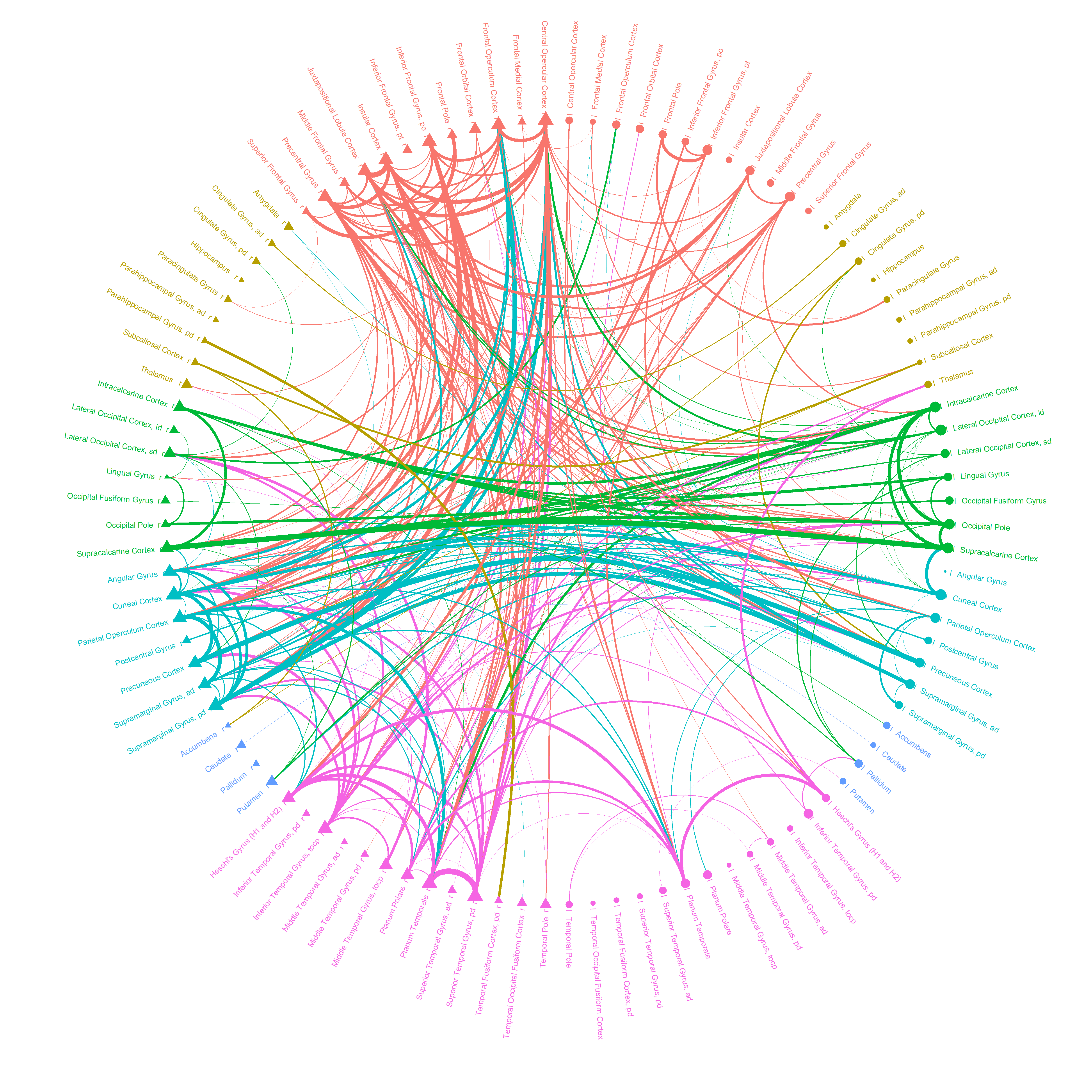}
%\hspace{1cm} 
\includegraphics[width=0.49\linewidth]{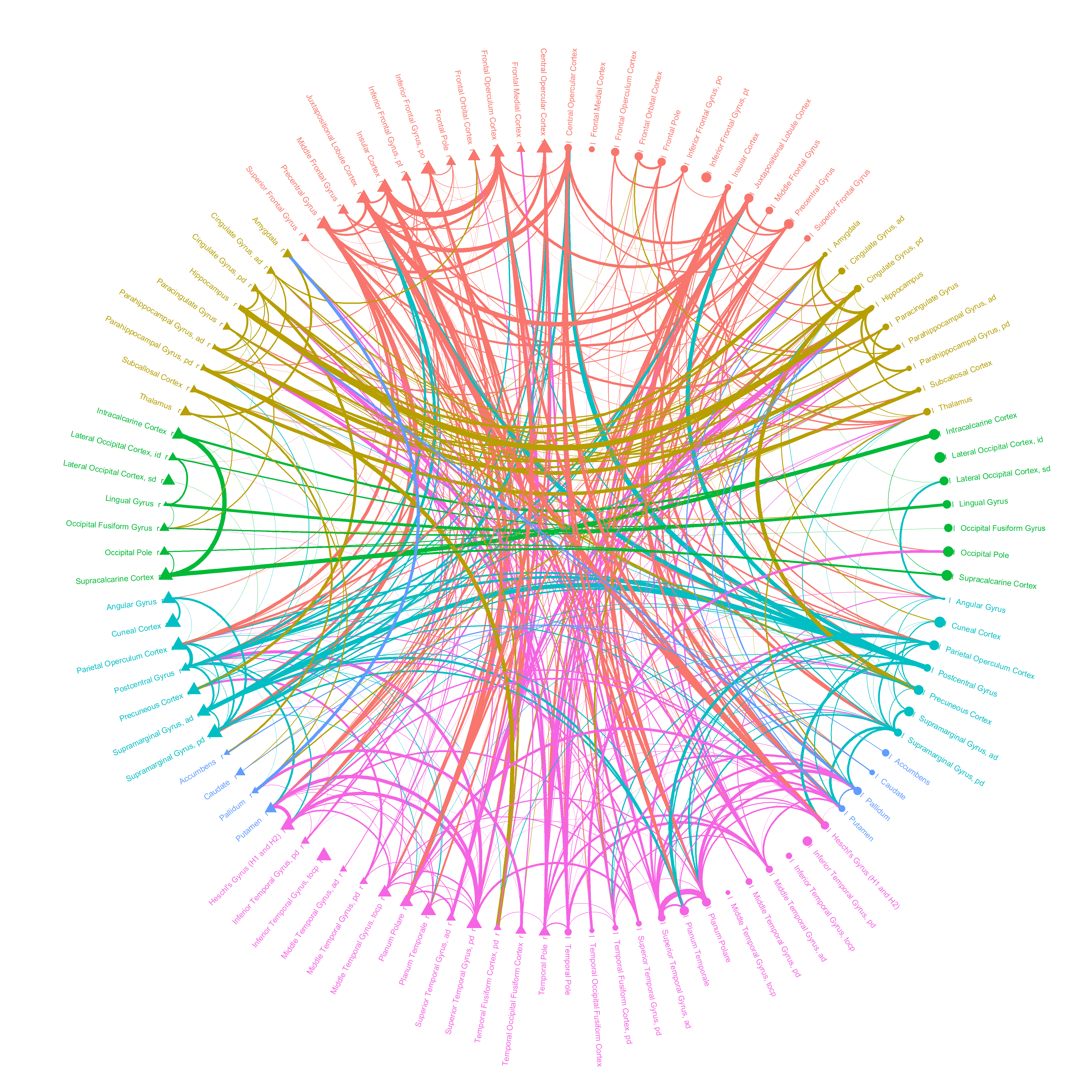}

\caption{Schematic view of information parity before and after ayahuasca ingestion. The illustration shows the information parity relations before (left) and after (right) Ayahuasca for one of the networks (mean degree $\langle k\rangle=28$) of subject 1. Each weighted link represents the information parity between the pair of brain regions. In order to improve the visualization, it depicted only $\rm IP \ge 0.25$.  The left and right hemispheres are marked by triangles and circles, respectively. The increase in information parity can be clearly observed. As there is a wider variation across subjects, we recommend that the reader does not draw conclusions based on observations of particular links.}
\label{fig:network}
\end{figure}

A schematic view of information parity increase can be observed in Fig.~\ref{fig:network}. It illustrates the information parity between the brain regions before and after Ayahuasca intake for one of the subjects (calculated from networks with same density, $\langle k\rangle=28$). The dots represent the brain regions, and the links represent the information parity between the nodes. One can observe clear differences on the information parity patterns. Although the average information parity increases for all subjects, the patterns of changes vary considerable for each subject. The same comparison for the other subjects are depicted in the supplementary material. 

\begin{figure}[ht]
\centering
a)\includegraphics[width=0.47\linewidth]{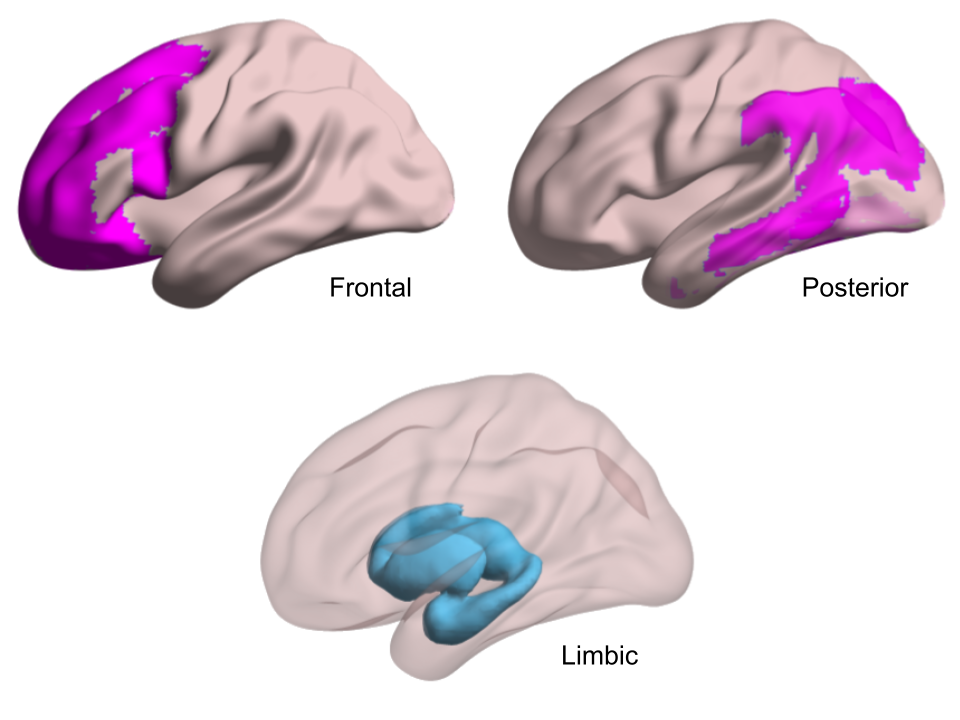}
b) \includegraphics[width=0.47\linewidth]{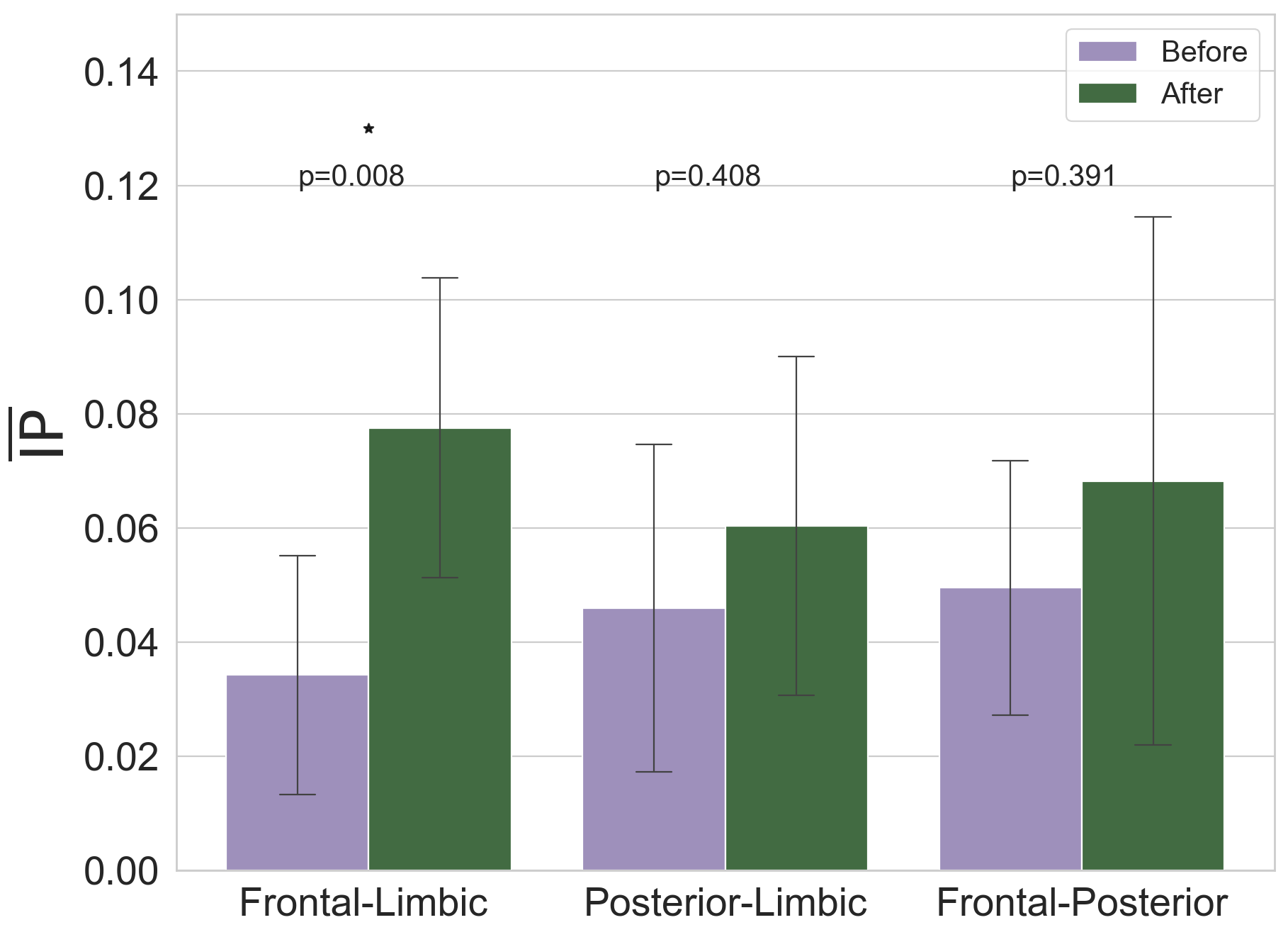}

c) \includegraphics[width=0.47\linewidth]{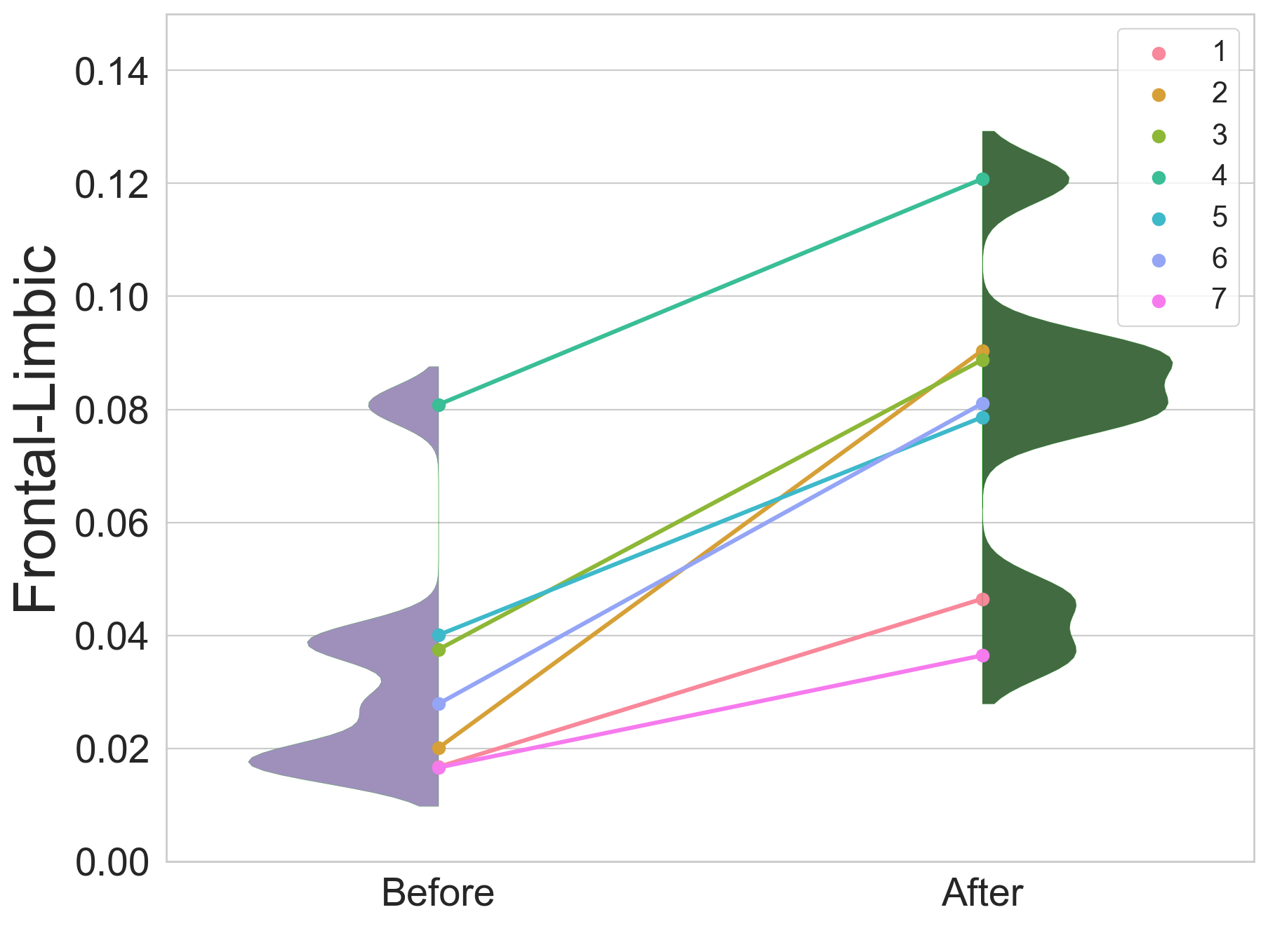}
 \includegraphics[width=0.47\linewidth]{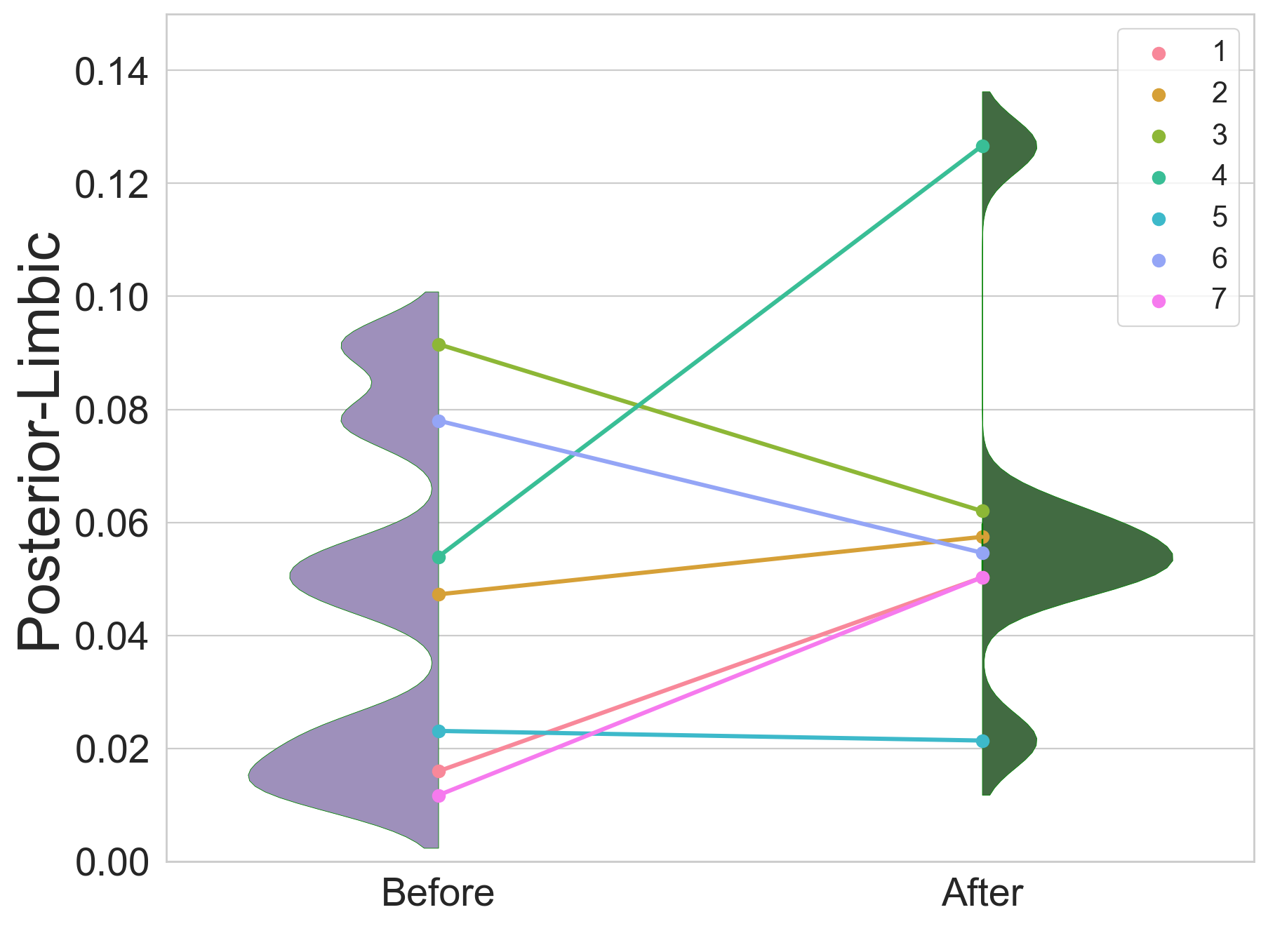}
\caption{Information parity between limbic regions and frontal and posterior brain regions. Panel a) shows a illustration of where the three clusters of regions are localized. The regions are listed in the table \ref{table:tab_listregions}. Panel b) compares the  average information parity across all subjects before (purple) and after (green) Ayahuasca intake ($k=28$) between limbic and frontal, limbic and posterior, and frontal and posterior regions. Note that only the frontal-limbic has a significant difference. In fact, only this regions have an increased information parity for all the subjects as can be observed in panel~c).    }
\label{fig:regional_comp}
\end{figure}

Next, we investigate if the information parity between large set of the brain regions would change consistently under the influence of Ayahuasca.
Guided by the literature, we focus our study on the relations between regions from posterior and frontal cortex  \cite{WAR13}  with regions of limbic system. The limbic structure considered here includes hippocampus, amygdalas and other regions that could play a role in memory retrieval psychedelic experiences. A sketch of the brain regions locations are presented in colorful shadows in Fig.~\ref{fig:regional_comp}~a). The precise regions are listed in Tab.~\ref{table:tab_listregions}.

In the bar plot of Fig.~\ref{fig:regional_comp}~b), we compare the average information parity between the limbic and frontal, limbic and posterior, and frontal and posterior regions. The latter is considered to serve as a control. The only significant and consistent changes for all individuals are the increase of information parity between frontal and limbic regions. The details of the changes for each individual can be observed in Fig.~\ref{fig:regional_comp}~b) and c). Note that the information parity between the limbic and frontal regions increases for all subjects, while the information parity between the limbic and posterior regions may increase or decrease, depending on the subject.   

\newpage

\section*{Discussion}

Humans have at all times used psychedelics to deliberately manipulate the state of consciousness. Currently, psychedelic substances had assume a central importance in neuroscience not only for its therapeutic potential \cite{PALF19,SAN16a,CAR18}, but also because it allow us explore the neural correlate of consciousness. 
Psychedelics substances affect the brain functions in a non straightforward way, therefore it is needed discerning analytical tools in order to understand it. 

Recent researches have been shown that under psychedelic influence the brain functions become less constrained as indicated by increasing entropy in different aspects of brain activity \cite{FEL21,TAG14,VIO17a,VIOL19,CAR19a}.
Along the increase on entropy, previous results have indicated an increase in network segregation after Ayahuasca intake characterized by an increase of local efficiency, clustering coefficient, and geodesic distances and decrease of global efficiency \cite{VIO17a}. Similar results was reported for individuals under the psychedelic substance Lysergic acid diethylamide (LSD) effects \cite{LUPPI21}.
Together, they suggest that, on average, the brain regions tends to have a less restrictive influence from overall brain network.
Indeed, the average geodesic entropy, that is, the entropy of the individual topological profiles of the nodes, is higher after Ayahuasca intake \cite{VIOL19}.
In this context, the increase in the average information parity could indicate that, although the overall influence on brain regions' activity tends to be less constrained, there is an increase in the redundancy of pairwise topological profiles. In other words, concerning the the topological symmetry, the functional influences shared by pair of regions are more similar.
Based on the results reported here, we hypothesizes that the release on the constraints does not results in an increase of disorder, instead, the functional networks reorganize itself in a way that the disruption in the hierarchy \cite{VIO17a} does not affect its resilience, since, at least, the pairwise informational redundancy increases. 

The cluster of regions defined here as limbic system is traditionally associated with processes as learning, emotions, memory, reward mechanisms, impulses control, among others.  The frontal cortex have been associated to top-down modulation \cite{BUS07}, what consist in a  high level processing that modulate low level processing such memory retrieval \cite{TOM99}, emotions control \cite{ERK07} to cite some. Furthermore, it has been recently hypothesized that top-down control is reduced under Ayahuasca influence, due a reduction of transfer entropy from frontal regions to other brain regions \cite{ALO15}. According to the authors, the reduction of constraints from frontal regions under psychedelic influences allow an increase of bottom-up information flow, characterized by a increased influence of low-level sensory cortices \cite{ALO15}. 
The increase in the information parity between their frontal cortex and limbic regions corroborate with the hypothesis of functional hierarchy disruption under psychedelic influences \cite{ALO15}. In fact, our results suggest that, from a statistical point of view, the frontal cortex and limbic system have a more comparable influence on the whole brain in the psychedelic state than in the normal state. 

The results report here can greatly add to the framework of knowledge about effect of psychedelics on brain and corroborate with literature hypothesis. However, some limitations have to be considered: (i) the number of individuals are small; (ii) our experimental design did not include a placebo group; (iii) for the local analysis, we choose the clusters of regions based on neuroscience literature instead perform a systematic  modular inspection. Hence, we strongly recommend a replication of the analysis performed here, alongside to the previous analysis \cite{VIO17a,VIOL19,FEL21}, for a large group of individuals and for other elevated or task-invoked states of mind as well. We deem that this framework can yield significant insights to the understand of the neural basis of consciousness. 

\begin{center}
\begin{table}[ht!]
\begin{tabular}{|l|l|}
\hline
Frontal Cortex & Frontal Pole R/L \\ %\cline{2-2} 
 & Superior Frontal Gyrus R/L \\ %\cline{2-2} 
 & Middle Frontal Gyrus R/L \\ %\cline{2-2} 
 & Inferior Frontal Gyrus pars opercularis R/L\\%\cline{2-2} 
 & Frontal region \ Juxtapositional Lobule Cortex R/L\\ %\cline{2-2} 
 & Paracingulate Gyrus R/L\\ %\cline{2-2}
 & Cingulate Gyrus, anterior division R/L\\ %\cline{2-2} 
 & Frontal Orbital Cortex R/L\\ %\cline{2-2} 
& Frontal Operculum Cortex R/L\\ \hline
Posterior Cortex & Middle Temporal Gyrus, posterior division R/L\\ %\cline{2-2} 
 & Middle Temporal Gyrus, temporooccipital part R/L\\ %\cline{2-2} 
 & Supramarginal Gyrus, anterior division R/L\\ %\cline{2-2} 
 & Supramarginal Gyrus, posterior division R/L\\ %\cline{2-2} 
 & Angular Gyrus R/L\\ %\cline{2-2} 
 & Lateral Occipital Cortex, superior division R/L \\ %\cline{2-2} 
 & Lingual Gyrus R/L\\ %\cline{2-2} 
 & Temporal Fusiform Cortex, posterior division R/L\\ %\cline{2-2} 
 & Temporal Occipital Fusiform Cortex R/L\\ %\cline{2-2} 
 & Occipital Fusiform Gyrus R/L\\ %\cline{2-2} 
 & Parahippocampal Gyrus, anterior division R/L\\ 
 \hline
Limbic System & Thalamus R/L\\ %\cline{2-2} 
 &  Caudate R/L\\ %\cline{2-2} 
 &  Putamen R/L\\ %\cline{2-2} 
 &  Pallidum R/L\\ %\cline{2-2} 
 &  Hippocampus R/L\\ %\cline{2-2}
 & Amygdala R/L\\
 &  Accumbens R/L\\ \hline
\end{tabular}
\caption{List of regions considered the 3 clusters we compared. The regions was defined according to the Harvard-Oxford cortical and subcortical structural atlas (threshold of $\ge 25 \%$, using FMRIB Software Library, www.fmrib.ox.ac.uk/fsl). }
\label{table:tab_listregions}
\end{table}
\end{center}

%\blankpage
\newpage

\bibliographystyle{naturemag-doi}
\bibliography{main.bib}

\end{document}